\documentclass[doublecol]{epl2}

\usepackage{color}
\usepackage{calc}
\usepackage{array}
\usepackage{graphicx}
\usepackage{amsmath, amssymb}
\usepackage{array}

\def\abs#1{\left \vert #1 \right \vert}
\def\op#1{\mathbf{#1}}

\title{Geometric scaling in the spectrum of an electron \\
       captured by a stationary finite dipole}
\shorttitle{Geometric scaling in the spectrum of a finite dipole}

\author{D{\'{a}}niel Schumayer\inst{1} \and
        Brandon P. van Zyl\inst{2}     \and
        Rajat K. Bhaduri\inst{3}       \and
        David A. W. Hutchinson\inst{1}}
\shortauthor{D. Schumayer \etal}

\institute{%
  \inst{1} Jack Dodd Centre for Quantum Technology,
           Department of Physics, \\
           University of Otago,
           730 Cumberland St,
           Dunedin 9016,
           New Zealand\\
  \inst{2} Department of Physics,
           St. Francis Xavier University, \\
           Antigonish, Nova Scotia B2G 2W5,
           Canada \\
  \inst{3} Department of Physics \& Astronomy,
           McMaster University,\\
           1280 Main St. West,
           Hamilton, Ontario K2H 4C3,
           Canada
}

\abstract{%
          We examine the energy spectrum of a charged particle in the presence
          of a {\it non-rotating} finite  electric dipole. For {\emph{any}}
          value of the dipole moment $p$ above a certain critical value
          $p_{\mathrm{c}}$ an infinite series of bound states arises of which
          the energy eigenvalues obey an Efimov-like geometric scaling law
          with an accumulation point at zero energy. These properties are
          largely destroyed in a realistic situation when rotations are
          included. Nevertheless, our analysis of the idealised case is of
          interest because it may possibly be realised using quantum dots
          as artificial atoms.
         }

\begin{document}

\maketitle

\section{Introduction}
The problem of a charged particle in the field of a physical electric
dipole serves as the starting point for the description of a variety
of important physical processes, ranging from e.g., the passage of
muons through a substance~\cite{Fermi1947}, the determination of
carrier mobilities and charge trapping states in various condensed
matter systems~\cite{Klahn1998, Fry2000}, to the chemical bonding of
two atoms~\cite{Morse1929, Desfrancois2004}. Despite the vast amount
of literature on the quantum mechanics of this system, there have
been remarkably few detailed, numerically exact studies~\cite{Wallis1960,
Power1973}. Past efforts have focused on the narrow region near
criticality where scattering states from the continuum are brought
down into the discrete spectrum~\cite{Crawford1967, Turner1968b,
Mezei2006, Papp2005}. The primary impetus for these studies is rooted
in chemistry where the determination of the critical dipole
moment(s), along with the binding energy of the charged particle, are
very important. It is well established that a non-rotating,
rigid dipole can only bind a charged particle if the dipole moment
exceeds the critical $p_{\mathrm{c}} \approx 1.6249$\,Debye.

While part of the reason for the heretofore limited numerical results
can be attributed to the computationally intensive nature of the
calculations, the more significant issue has likely been the absence
of additional physical motivation to warrant further studies.
It is now known that the inclusion of rotation drastically
alters the idealised spectrum\cite{Garrett1971, Garrett1980} of the
stationary dipole. With the advent of quantum dots as artificial
atoms, however, there is a possibility that the rotational degrees of
freedom are irrelevant. In that case the idealised model of this
paper may be of interest to the ultra-cold atoms community, who
look for such a spectrum at Efimov resonance \cite{Knoop2009,
Ferlaino2009}.

In this Letter, we present a detailed, numerically exact investigation
for the bound-state energy spectrum of a charged particle in the field
of a stationary finite dipole. This model occurred in the physics
literature as early as 1947~\cite{Fermi1947}. It was firmly
established and later ``rediscovered''~\cite{Turner1977} that a finite
stationary dipole can capture an electron if the dipole moment, $p$,
is larger than a critical value $p_{c}$. Furthermore, for supercritical
values of $p$ not only one but infinitely many bound state appears in
the spectrum, with an accumulation point at $E=0$.

We focus our attention on the same system: a rigid, stationary
dipole, hence we exclude the effect of rotation of the dipole from
our calculation. Nevertheless we have to note that rotation may play
an important role in influencing the spectrum of the electron,
especially if the moment of inertia, $I$, of the dipole is small.
Garrett~\cite{Garrett1971, Garrett1980} showed that the critical
dipole moment, $p_{c}$, is larger compared to that of a stationary
dipole and depends on $I$. Additionally, for $p > p_{c}$ the number
of bound states is reduced to a finite value and the infinitude of
the number of bound states is only recovered if $I \rightarrow
\infty$.

Until now the striking similarity between the detailed spectrum of
our system and the three-body Efimov levels \cite{Efimov1970,
Braaten2007} has not been emphasised. Despite the different origins
of these two effects~\cite{Amado1971}, our numerical and analytical
results of the bound electronic spectrum for $p>p_c$ yield two
features that resemble an Efimov-spectrum:
\begin{itemize}
   \item{a geometric scaling law of consecutive binding energies,
         $E_{n+1}/E_{n}= {\mathrm{constant}}$, and}
   \item{an accumulation of the bound states as $E \rightarrow 0^{-}$.}
\end{itemize}
\begin{figure}[bht!]
   \centering\includegraphics[height=45mm]{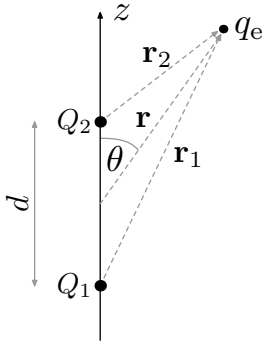}
   \caption{\label{fig:model}%
            Schematic representation of the model. The line joining
            the center of the dipole to $q_e$ is of length $r$, and
            makes an angle $\theta$ with the $z$-axis.}
\end{figure}

\section{The model} Our physical model consists of three charged
particles as schematically illustrated in figure~\ref{fig:model}.
We keep the position of two particles fixed (i.e., $Q_{1}$ and $Q_{2}$)
and look for the bound state(s), if such exists, of the third
particle. Specifically, for the stationary rigid dipole, we choose
$Q_{1} = -Q_{2} = Q$, and the third particle is taken to be an
electron. Therefore the Hamiltonian, $\op{H}$, for our system reads
\begin{equation} \label{eq:PhysicalHamiltonian}
   \op{H}
   = - \frac{\hbar^2}{2m_e} \nabla^{2}
     + \frac{Qq_e}{4\pi \epsilon_{0}}
       \left (
              \frac{1}{\abs{ {\mathbf{r}}_{1} }}
              -
              \frac{1}{\abs{ {\mathbf{r}}_{2} }}
       \right )~.
\end{equation}
It is convenient to introduce the dimensionless dipole moment,
$\lambda$, with the definition
\begin{equation}
   \lambda = \frac{2 m_{\mathrm{e}}}{\hbar^{2}} \ 
             \frac{pq_{\mathrm{e}}}{4\pi \epsilon_{0}}~,
\end{equation}
where $p$ denotes the electric dipole moment, $p=Qd$, and $d$ is
the size of the physical dipole. Finally, scaling all lengths as
${\mathbf{r}}_{i} \rightarrow {\mathbf{r}}_{i}/d$, reduces
eq.~(\ref{eq:PhysicalHamiltonian}) to
\begin{equation} \label{eq:ReducedHamiltonian}
   \op{H}
   = - \nabla^{2}
     + {\lambda}
       \left (
              \frac{1}{\abs{ {\mathbf{r}}_{1} }}
              -
              \frac{1}{\abs{ {\mathbf{r}}_{2} }}
       \right )~.
\end{equation}
In our simplified picture, $\lambda$ characterises the strength of
the interaction between the originally unbound electron and the
physical dipole. If $\lambda$ does not exceed the threshold value
$\lambda_c \approx 1.27863$, the third particle cannot be bound.
The minimum $\lambda$ then represents the critical strength at
which the first zero-energy bound state appears. Note, from eq.~
(\ref{eq:ReducedHamiltonian}), that for $r \gg d$, the potential
goes like $\cos{\!(\theta)}/r^{2}$ in spherical polar co-ordinates,
where $r$ and $\theta$ are defined in figure~\ref{fig:model}. For
$r \ll d$, however, the potential has no singularity, and goes to
zero as $r \rightarrow 0$. At threshold, the spatial extent of the
wavefunction for the third particle becomes much greater than the
size of the dipole~\cite{Chatterjee2008}, and the short-distance
length scale, $d$, set by the size of the dipole may be ignored. As
a result, the same critical value $\lambda_c$ is obtained for both
the finite and for the {\emph{point}} dipole~\cite{Turner1966,
Brown1967, Connolly2007, Alhaidari2008}. For $p>p_c$, however, the
point dipole potential $\cos{\!(\theta)}/r^{2}$ has no lower bound
in energy because of the $r^{-2}$ singularity at the
origin~\cite{Connolly2007}, and is unphysical. By contrast, our
finite-sized dipole potential is well-behaved for all ${\textbf{r}}$,
thereby yielding a meaningful bound state spectrum for $p>p_{c}$.
Moreover, it is the inverse-square nature of this potential for
large distances that gives rise to the Efimov-like features of
the spectrum~\cite{Efimov1981}.

{\emph{Numerical implementation}} --- The bound state energies,
$E<0$, and wavefunction, $\Psi$, for the electron are obtained
from the reduced Schr{\"o}dinger equation,
\begin{equation}
   \op{H} \Psi = - \kappa^{2} \Psi,
\end{equation}
where $\kappa^2 = -2m_{e} E d^{2}/\hbar^2 \ge 0$ is the
dimensionless energy. Equation (4) is separable in prolate
spheroidal coordinates~\cite{Judd1975}, resulting in the
following equations ($\varphi \in [0, 2\pi [$, $\xi \in [1,
\infty [$ and $\eta \in [-1, 1]$)
\begin{subequations} \label{eq:TheODEs}
\begin{align}
   \label{eq:ODEphi}
   \frac{d F}{d\varphi} + m^2 F = 0 \\
   \label{eq:GSExi}
   \frac{d }{d\xi}\!
      \left \lbrack
         (\xi^{2}-1) \frac{d S}{d\xi}
      \right \rbrack
   + \!\left \lbrack
        - A + \kappa^{2} \xi^{2} - \frac{m^{2}}{\xi^{2}-1}
     \right \rbrack S = 0 \\
   \label{eq:GSEeta}
   \hspace*{-4mm}
   \frac{d }{d\eta}\!
     \left \lbrack 
        (1-\eta^{2}) \frac{d T}{d\eta}
     \right \rbrack
   + \!\left \lbrack
        A - \lambda \eta - \kappa^{2} \eta^{2} - \frac{m^{2}}{1-\eta^{2}}
     \right \rbrack T = 0
\end{align}
\end{subequations}
where the wavefunction has been factorised as $\Psi = S(\xi) T(\eta)
F(\varphi)$. The reduction of the three-dimensional Schr{\"o}dinger
equation to a set of three one-dimensional ordinary differential
equations (ODEs) requires the appearance of two separation constants,
$A$ and $m$. Separability, as the manifestation of a symmetry, also
means here that these constants are the eigenvalues of conserved
quantities~\cite{Erikson1949, Coulson1967a}. If there is no electric
dipole, i.e. $\lambda=0$, the electron is free and the solution of
the original physical problem is known. In this case the two
conserved quantities are $\op{L}^{2}$ and $\op{L}_{z}$ with
eigenvalues $\ell(\ell+1)$ and $m$, where $\ell$=0,1,\dots and $m=$0,
$\pm$1, \dots $\pm \ell$. Since our potential does not possess
spherical symmetry for $\lambda > 0$, the ($2\ell+1$) degeneracy of
$m$ is absent, although the azimuthal symmetry of the finite dipole
still guarantees the conservation of $\op{L}_{z}$. Notice that the
same is not true for $\op{L}^{2}$, therefore $\ell$ is no longer a
good quantum number. In what follows, we adopt the notation
established previously in the literature~\cite{Wallis1960,
Abramov1972, Power1973} and label the energy eigenvalues of $\op{H}$
by a set of three numbers ($n_{\xi}$, $n_{\eta}$, $m$) where
$n_{\xi}$ counts the zeros of $S(\xi)$, while $n_{\eta}$ does
so for $T(\eta)$.

Although the general solution of the above equations can be
analytically given in terms of double-confluent Heun
functions~\cite{Ronveaux1995}, this approach does not offer a
feasible way for deriving the energy eigenvalues explicitly.
Consequently, we fall back to numerical methods~\cite{Bates1953,
Makarewicz1989} for finding normalisable solutions to these
equations for a given value of $\lambda$ and $\kappa^{2}$.

In the first instance, we assume that the separation constant $A$
can be different in the two equations. In this way we obtain two
sets of curves, viz.  $\kappa^2$ vs. $A_{\mathrm{rad}}$ and
$A_{\mathrm{ang}}$, for the radial (\ref{eq:GSExi}) and for the
angular equation (\ref{eq:GSEeta}), respectively. Only the latter
curve, $A_{\mathrm{ang}}$, depends on the dimensionless dipole
moment $\lambda$, since $\lambda$ is present in (\ref{eq:GSEeta}),
but absent in (\ref{eq:GSExi}). The crossing(s) of these two sets
of curves represents the desired solution to the system of ODEs
since the separation constant $A = A_{\mathrm{rad}} = A_{\mathrm{ang}}$
and $\kappa^{2}$ become common for the two equations.

\begin{figure}[thb!]
   \hspace*{-2mm}
   \includegraphics[angle=-90,width=1.01\textwidth/2-3mm]{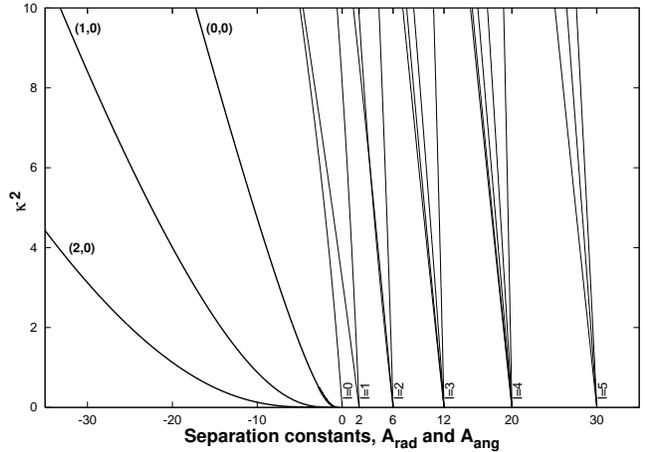}
   \caption{\label{fig:SeparationConstLambda0}
            The angular (thin lines) and radial (thick lines) separation constants
            (\ref{eq:GSExi}-c) are plotted versus the dimensionless
            energy eigenvalue $\kappa^{2}$ for $\lambda=0$. Each curve 
            of $A_{\mathrm{ang}}$, except the ones for $m=0$, is doubly
            degenerate, therefore each branch originating from $\ell (
            \ell+1)$ has ($2\ell+1$) curves altogether. The curves of
            the radial separation constant are labelled by ($n_{\xi}$,
            $m=0$). If $\lambda < \lambda_{\mathrm{c}}$ there is no
            crossing at positive $\kappa$, and there is no bound state.
           }
\end{figure}

\begin{figure}[thb!]
   \includegraphics[angle=-90,width=\textwidth/2-3mm]{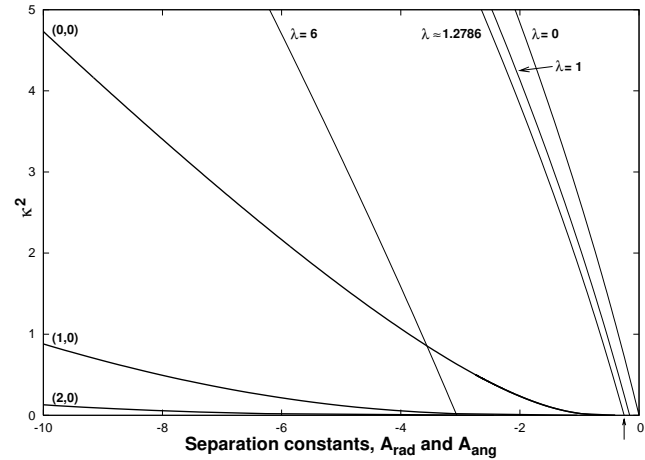}
   \caption{\label{fig:SeparationConst}
            The angular (thin line) and radial (thick line)
            separation constants are plotted versus the dimensionless
            energy eigenvalue $\kappa^{2}$. The curves of the radial
            separation constant are labelled by ($n_{\xi}$, $m=0$).
            Their rapid decrease as energy approaches zero is
            apparent. Moreover, figure shows only ($n_{\eta}=0$,
            $m=0$) angular separation constant for four different
            $\lambda$ values: two below the critical value, one at
            the critical value ($\approx$ 1.2786) and one much larger
            than the critical value. The small vertical arrow at
            $A_{\mathrm{rad}} = A_{\mathrm{ang}}=-1/4$ captures the
            position where the zero-energy bound states first appear.
           }
\end{figure}

Figure \ref{fig:SeparationConstLambda0} shows the two sets of curves,
$A_{\mathrm{rad}}$ and $A_{\mathrm{ang}}$ for $\lambda \equiv 0$,
which case represents a free electron, therefore $\op{L}^{2}$ is
conserved and $\ell$ is a good quantum number to label the curves
with. The fan structure of $A_{\mathrm{ang}}$ is apparent in the
figure and remains intact for non-vanishing dipole moments. For $0 <
\lambda < \lambda_{\mathrm{c}}$, there is no common point for the
two sets of curves in the $\kappa^{2}>0$ domain, thus no bound state
exists. On one hand, it can also be proven, that $\lim_{\kappa^{2}=
0^{+}} {\left ( A_{\mathrm{rad}} \right )} = -1/4$, i.e. all
$A_{\mathrm{rad}}$ curves approach the value ($-1/4$). On the other hand
as we increase the dipole moment the $A_{\mathrm{ang}}$ curves shift
towards left (see fig.~\ref{fig:SeparationConst}). Therefore there
must be a critical value of the dimensionless dipole moment $\lambda$
at which the first $A_{\mathrm{ang}}$ curve, labelled by ($n_{\eta}=
0, m=0$), also crosses the abscissa at ($-1/4$). However, above this
critical value of $\lambda$, $A_{\mathrm{ang}}$(0,0) crosses
infinitely many curves of $A_{\mathrm{rad}}$; this is the
mathematical origin of the infinite ``tower'' of bound states if
$\lambda \geq \lambda_{\mathrm{c}}$. Figure \ref{fig:SeparationConst}
depicts $A_{\mathrm{ang}}(n_{\eta}, m)$ only for $(n_{\eta}=0, m=0)$.
Increasing $\lambda$ further, more and more branches of the ``fan''
are shifted to the left and cross the value ($-1/4$), hence one can
deduce a critical value of $\lambda$ for each ``tower'' labelled by
$(n_{\eta}, m)$. The value of the five lowest $\lambda_{c}(n_{\eta},
m)$ are reported in table \ref{table:CriticalLambda}. As $\lambda$
exceeds a given $\lambda_{c}(n_{\eta}, m)$ a new ``tower'' of infinitely
many bound states opens, e.g. for $\lambda = 16$ there will be three
sets, ($n_{\eta}, m$) = (0,0), (0,1) and (1,0) each containing
infinitely many bound states labelled by $n_{\xi}=$0, 1, 2, \dots

The values of $A_{\mathrm{ang}}$ and $A_{\mathrm{rad}}$ are
calculated as eigenvalues of matrices or from three-term
recursions~\cite{Bates1953, Makarewicz1989}. Numerically,
it is demanding to obtain precise values for the crossings,
particularly for the rapidly decaying $A_{\mathrm{rad}}$ curves near
the critical region $\lambda \rightarrow \lambda_{\mathrm{c}}^{+}$.

\section{Results} For a given set of $(n_{\eta},m)$, if $\lambda
\geq \lambda_{\mathrm{c}}(n_{\eta},m)$ there are an infinite tower of
bound states, irrespective of the value of $\lambda$. This result is
in marked contrast to what is found in the Efimov effect, where the
infinite tower of bound states for the three-body system occurs only
at criticality. We note that the lowest critical dipole moment
$\lambda_{\mathrm{c}} \approx 1.2786$ corresponds to an electron with
quantum numbers (0,0,0), that gives a zero-energy bound state. For
$n_{\eta}>0$, it is well-known in the literature that higher values
of $\lambda_c$ are needed. Our exact numerical results (see Table
\ref{table:CriticalLambda}) confirm these super-critical
values~\cite{Alhaidari2008} with high accuracy.
\begin{table}[bht!]
   \caption{\label{table:CriticalLambda}
            The five lowest values of $\lambda_{\mathrm{c}}
            (n_{\eta}, m)$ are given. If $\lambda >
            \lambda_{\mathrm{c}}(n_{\eta}, m)$ a new set of infinite
            set of bound states is created.
           }
   \centering
   \begin{tabular}{p{5mm}p{5mm}rcr}
      $n_{\eta}$ & $m$ &  $\lambda_{\mathrm{c}}$ \hspace*{4mm} & &  $p_{\mathrm{c}}$ (Debye) \\
      \hline\hline
          0     &   0  &  1.2786298 & &  1.625 \\
          0     &   1  &  7.5839359 & &  9.638 \\
          1     &   0  & 15.0939114 & & 19.182 \\
          0     &   2  & 19.0580547 & & 24.220 \\
          1     &   1  & 28.2242292 & & 35.869 \\
      \hline\hline
   \end{tabular}
\end{table}

\begin{figure}
   \includegraphics[angle=-90,width=\textwidth/2-3mm]{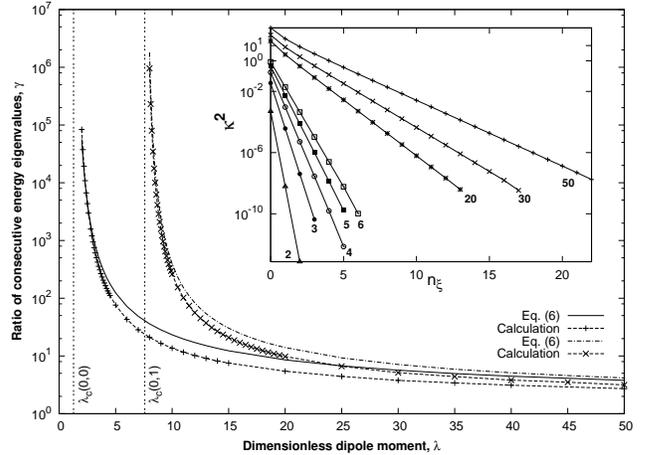}
   \caption{\label{fig:RatioOfEnergyEigenvalues}
            The ratios of consecutive energy eigenvalues, $\gamma$,
            are plotted for $m=0$ and $m=1$ in a semi-logarithmic
            graph. The solid thin and dash-dotted lines represent
            the analytical approximation \eqref{eq:RatioAnalytical}
            while the calculated data are represented by $+$ ($m=0$)
            and by $\times$ ($m=1$). The dotted vertical lines depict
            the $\lambda_{\mathrm{c}}$ values (see table
            \ref{table:CriticalLambda}). Inset: dimensionless
            energy eigenvalue $\kappa^{2}$ as a function of $n_{\xi}$
            for different values of $\lambda$ in a semi-logarithmic
            graph. Curves are labelled with increasing values of
            $\lambda$ from left-to-right. Lines are a guide to the
            eye.
            }
\end{figure}

Figure~\ref{fig:RatioOfEnergyEigenvalues} summarises the central
results of this paper. Focusing first on the curves with symbols,
the main figure shows the ratio of consecutive bound state energies
for a variety of dimensionless dipole moments. It is evident that at
criticality, $\lambda = \lambda_{\mathrm{c}}$, the ratio diverges,
which is expected given that this is the dipole moment for which the
{\emph{zero energy}} state appears. In addition, we see that above
criticality, the ratio $\gamma$, decays exponentially. Similar
curves are also found for other values of $\lambda_{\mathrm{c}}
(n_{\eta},m)$, and when examined in detail, these numerical findings
suggest a universal form for $\gamma$ near criticality. Indeed,
following the earlier work of Abramov~\cite{Abramov1972}, one can
deduce a simple, universal expression for the ratio of successive
eigenvalues valid near the critical region ($n_{\xi} \ge 0$,
$n_{\eta}$ and $m$ are fixed)
\begin{equation} \label{eq:RatioAnalytical}
   \gamma \cong
   \frac{E(n_{\xi}+1, n_{\eta}, m)}%
        {E(n_{\xi}, n_{\eta}, m)}
   =
   \exp{
        \!\left (
           \frac{\Gamma^{2} \!\left ( \frac{1}{4} \right )}%
                {\sqrt{2 (\lambda - \lambda_{\mathrm{c}}(n_{\eta}, m))}}
        \right )
       }
\end{equation}
The $\gamma$ values obtained from eq.~(\ref{eq:RatioAnalytical}) are
also shown in the main panel of figure~\ref{fig:RatioOfEnergyEigenvalues}.
Note that very near criticality, the analytical and numerical ratios
are indistinguishable. However, as we leave the critical region, the
two curves begin to deviate appreciably, with the result that the
analytical prediction overestimates $\gamma$. The linear relationship
between the $\kappa^2$ and $n_{\xi}$ in the inset to
fig.~\ref{fig:RatioOfEnergyEigenvalues} clearly establishes the
Efimov-like scaling for the bound state energies.

\begin{figure}[bht!]
   \includegraphics[width=0.47\textwidth]{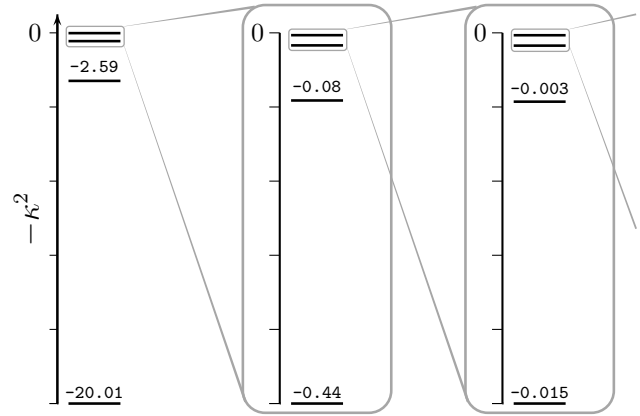}
   \caption{\label{fig:MatryoshkaScaling}
            Dimensionless energy eigenvalues $\kappa^{2}$ of the (0,0)
            band are shown for $\lambda=20$. Accumulation of energy
            levels as $(-\kappa^{2}) \rightarrow 0^{-}$ is obvious and
            also demonstrated by magnifying the spectrum around the
            zero energy. The scales on each axis are different.
           }
\end{figure}

In fig.~\ref{fig:MatryoshkaScaling}, we display the dimensionless
energy spectrum of the electron in the field of the finite electric
dipole for $\lambda=20$, for which $\gamma \approx 5.38$. The
accumulation of bound states as $E \rightarrow 0^{-}$ is emphasized by
magnifying the spectrum around zero energy, and continues indefinitely.

\section{Outlook} We propose an experimentally feasible scenario to
observe this geometrical scaling. Our suggestion involves positively
and negatively charged quantum dots~\cite{Findeis2001}. By assembling
two such dots (representing our fixed charges) onto an insulator
matrix one could design a near ideal representation of our model
system. This realisation lacks the full rotational symmetry around
the axis of the dipole, contrary to our model analysed above, however
it only restricts the possible values of the $m$ quantum number to
even integers and should not alter the infinite number of bound
states and their scaling property. The loosely bound states could be
detected by standard photo-detachment spectroscopy~\cite{Mead1984}.
The additional advantage of this scenario would be the possibility
of tailoring the dipole moment to a favourable value.

\section{Conclusion} We have investigated the energy spectrum of
a charged particle's bound states in the field of a fixed, finite
dipole. Using an exact numerical approach, we have confirmed the
numerical values of previously known critical dipole moments required
to bind the charged particle in different symmetry states. The new
result is that for each set of fixed quantum numbers ($n_{\eta}$,
$m$) there are an infinite number of bound states, labelled by
$n_{\xi}$, for which the consecutive energy eigenvalues are ordered
in geometrical scaling. We refer to this scaling as
{\emph{Efimov-like}} because geometric scaling of the bound state
energy levels is a signature of the Efimov effect. We also find an
accumulation of states near zero energy, as in the Efimov states.
However, our system is not equivalent to Efimov's original scenario.
In the classic Efimov effect, three identical (neutral) bosons
interact pair-wise {\it via} a short-range potential, with the
result that an infinite series of excited {\em three-body} energy
levels appear {\em{only}} when at least two of the two-body
subsystems are at threshold. In our model, we have a long-ranged
electrostatic dipolar interaction, and the three particles are
distinguishable. Furthermore, the geometric scaling and accumulation
of energies, $E\rightarrow 0^{-}$, in our system occurs for {\em any}
$\lambda \geq \lambda(n_{\eta},m)$. Nevertheless, the common origin
for the geometric scaling in both cases is the presence of an inverse
square potential in the description of the bound state energy
spectrum of the system. We have also suggested a possible
experimental scenario in which the results of this paper could
be investigated.

\acknowledgments
This work was supported under contract NERF-UOOX0703 (NZ) and also
by the University of Otago. BPvZ and RKB acknowledge financial
support from the Natural Sciences and Engineering Research Council
(NSERC) of Canada. DS is grateful to George H. Rawitscher and
J{\'o}zsef Fort{\'a}gh for discussions.


\end{document}